\documentclass{svjour3}                     
\smartqed  
\usepackage{graphicx}
\usepackage{amssymb}
\usepackage{amsmath}
\journalname{Few-Body Systems (FB20)}
\makeatletter
\def\simleq{\mathrel{\mathpalette\gl@align<}}
\def\simgeq{\mathrel{\mathpalette\gl@align>}}
\def\gl@align#1#2{\lower.6ex\vbox{\baselineskip\z@skip\lineskip\z@
     \ialign{$\m@th#1\hfill##\hfil$\crcr#2\crcr\sim\crcr}}}
\makeatother

\newcommand{\bec}[1]{\mbox{\boldmath $#1$}}

\newcommand{\Pu}{p_{\uparrow}}

\newcommand{\Nu}{n_{\uparrow}}
\newcommand{\Nd}{n_{\downarrow}}

\begin{document}

\title{
Few-baryon interactions from lattice QCD
}


\author{
Takumi Doi 
for HAL QCD Collaboration
}



\institute{T. Doi \at
Theoretical Research Division, Nishina Center, RIKEN, Wako 351-0198, Japan \\
              \email{doi@ribf.riken.jp}           
}

\date{Received: date / Accepted: date}

\maketitle

\begin{abstract}
We report the recent progress on
the determination of three-nucleon forces (3NF) in lattice QCD.
We utilize the Nambu-Bethe-Salpeter (NBS) wave function
to define the potential in quantum field theory, and 
extract two-nucleon forces (2NF) and 3NF on equal footing.
The enormous computational cost for 
calculating multi-baryon correlators on the lattice
is drastically reduced 
by developing a novel contraction algorithm
(the unified contraction algorithm).
Quantum numbers of the three-nucleon (3N) system are chosen to be
$(I, J^P)=(1/2,1/2^+)$ (the triton channel),
and we extract 3NF in which three nucleons are aligned linearly with an equal spacing.
Lattice QCD simulations are performed using
$N_f=2$ dynamical clover fermion configurations
at the lattice spacing of $a = 0.156$ fm on a $16^3\times 32$ lattice
with a large quark mass corresponding to $m_\pi= 1.13$ GeV.
Repulsive 3NF is found at short distance.

\end{abstract}

\section{Introduction}
\label{intro}

Nuclear (and baryonic) forces are 
fundamental quantities in nuclear physics,
and it is of crucial importance to determine them
from the underlying 
theory, 
Quantum Chromodynamics (QCD).
In particular, the determination of few-baryon forces
has a huge impact on
not only nuclear physics but also astrophysics.

There are various phenomena where
three-nucleon forces (3NF) are considered to play 
an important role, e.g., 
the binding energies of light nuclei~\cite{Pieper:2007ax},
the properties of neutron-rich nuclei and the supernova nucleosynthesis~\cite{Otsuka:2009cs}
and
the nuclear equation of state (EoS) as well as 
the saturation point of nuclear matter~\cite{Akmal:1998cf}.
Deuteron-proton  elastic scattering experiments 
have also shown a clear indication of 3NF~\cite{Sekiguchi:2011ku,Sekiguchi:FB20}.
Recent observational data on the maximum mass of neutron stars~\cite{Demorest:2010bx} 
triggered renewed interest in the nuclear EoS at high density,
and universal short-range repulsion for three baryons (nucleons and hyperons) 
may be needed 
in neutron stars with hyperon core~\cite{Nishizaki:2002ih,Takatsuka:2008zz}.

Despite of its phenomenological importance,
microscopic understanding  of 3NF is  still limited.
Pioneered by Fujita and Miyazawa~\cite{Fujita:1957zz},
the long range part of 3NF has been commonly modeled
by the two-pion exchange (2$\pi$E),
particularly with the $\Delta$-resonance. 
This 2$\pi$E-3NF 
is known to have 
an attractive nature at long distance,
and is considered to play a dominant role to resolve the problem
that light nuclei are underbound with 2NF only.
Unfortunately, the 2$\pi$E-3NF component alone cannot represent
the whole properties of 3NF:
For instance,
it would provide too strong attraction,
and thus incurs the overbound problem for light nuclei.
In order to handle this issue,
an additional repulsive component of 3NF at short distance is 
often introduced.
In Tucson-Melbourne model~\cite{Coon:2001pv},
a phenomenological cut-off parameter, $\Lambda \simeq 0.7$ GeV, is introduced 
in the (monopole) form factor of $\pi NN$ vertex~\cite{Nogga:1997mr}.
Compared to $\Lambda \simeq 1.3$ GeV 
obtained from $NN$ scattering data~\cite{Machleidt:1987hj},
one observes that short range repulsion is implicitly added in 3NF.
In Urbana/Illinois models, on the other hand,
a repulsive 3NF component at short distance is explicitly introduced
in a purely phenomenological way~\cite{Pieper:2001ap}.
Recently,
an approach based on the chiral effective field theory ($\chi$EFT)
is also developing~\cite{Machleidt:2011zz,Machleidt:2012xz},
and found to be useful to classify and parametrize 
the two-, three- and more-nucleon forces.
The $\chi$EFT-3NF contains not only pion exchange terms but also
short range terms.
The LECs ($c_D, c_E$) corresponding to the latter 
are known to be sensitive to the cut-off parameter 
in $\chi$EFT~\cite{Epelbaum:2002vt,Skibinski:2011vi}.
We note that repulsive short-range 3NF component
is phenomenologically required 
to explain the properties of high density matter~\cite{Akmal:1998cf,Nishizaki:2002ih,Takatsuka:2008zz}
and its detailed information is one of the central question of interest.

Since 3NF is originated by the fact that
a nucleon is not a fundamental particle,
it is most desirable to determine 3NF
directly from QCD,
in particular at short distance
where the dynamics of fundamental degrees of freedom (DoF), quarks and gluons,
becomes essential.
In this proceeding,
we report our progress on first-principles calculations of 
3NF using lattice QCD simulations~\cite{Doi:2010yh,Doi:2011gq}.
Note that 
while 
there are lattice QCD works for 
three- (or more) baryon systems%
~\cite{Yamazaki:2009ua,Yamazaki:2012hi,Beane:2009gs,Beane:2012vq},
they focus on the 
energies of the multi-baryon systems,
 and extracting 3NF
is currently beyond their scope.

As for the calculation of two-nucleon forces (2NF) from lattice QCD,
 an approach based on  the 
 NBS wave function 
 has been  proposed~\cite{Ishii:2006ec,Aoki:2008hh,Aoki:2009ji},
so that the potential is faithful to the phase shift
by construction. 
 Resultant (parity-even) 2NF 
are found to have desirable features
such as
attractive wells at long and medium
distances  and central repulsive cores at short distance.
The method has been  
applied to 
general hadronic interactions%
~\cite{Nemura:2008sp,Inoue:2010hs,Sasaki:2010bi,Inoue:2010es,Aoki:2011gt,Ikeda:2011qm,Murano:2011nz,Inoue:2011ai,Murano:2011aa}.
%
The method itself has been also generalized to ``time-dependent'' HAL QCD method,
so that the energy-independent (non-local) potential can be extracted 
without ground state saturation~\cite{HALQCD:2012aa}.
See Refs.~\cite{Aoki:2009ji,Aoki:2011ep,Aoki:2012tk} for recent reviews.
In this report, we extend the method to three-nucleon (3N) systems,
and perform the lattice QCD simulations for 3NF 
in the triton channel,
$(I, J^P)=(1/2,1/2^+)$~\cite{Doi:2010yh,Doi:2011gq}.
In particular, we update the results in Ref.~\cite{Doi:2011gq}
by employing the time-dependent HAL QCD method~\cite{HALQCD:2012aa}
to suppress the systematic error associated with excited states.
We also utilize a novel contraction algorithm (unified contraction algorithm), 
which reduces the enormous computational cost for 3N correlators
by a factor of 192~\cite{Doi:2012xd}.

\vspace*{-3mm}
\section{Formalism}
\label{sec:formulation}

We first 
briefly explain the framework for 
the calculation of 2N potentials~\cite{Aoki:2009ji,Aoki:2011ep,Aoki:2012tk}.
The central quantity in the HAL QCD method is the equal-time 
Nambu-Bethe-Salpeter (NBS) wave function,
$\phi_{2N}(\vec{r}) \equiv \langle 0 | N(\vec{r}) N(\vec{0}) | E_{2N}\rangle$,
where 
$|E_{2N}\rangle$ denotes the state of the 2N system at the energy of $E_{2N}$
in the center-of-mass frame,
$N$ the nucleon operator where spinor/flavor indices are implicit.
For simplicity, we consider the elastic region hereafter,
while the method can be extended above inelastic threshold~\cite{Aoki:2011gt,Aoki:E-indep:inelastic}.
%
The important property of the wave function is that
it has a desirable asymptotic behavior, 
\begin{eqnarray}
\phi_{2N} \propto \frac{\sin(kr-l\pi/2 + \delta_l^k)}{kr}, 
\quad 
r \equiv |\vec{r}| \rightarrow \infty,
\end{eqnarray}
where $E_{2N} = k^2/(2\mu)$ with the reduced mass $\mu = m_N/2$,
$l$ being the orbital angular momentum.
Exploiting this feature,
we define the (non-local) 2N potential, $U_{2N}(\vec{r},\vec{r}')$,
through the following Schr\"odinger equation,
\begin{eqnarray}
-\frac{\nabla^2}{2\mu} \phi_{2N}(\vec{r})
+ \int d\vec{r}' U_{2N}(\vec{r},\vec{r}') \phi_{2N}(\vec{r}')
= E_{2N} \phi_{2N}(\vec{r}) .
\label{eq:Sch_2N:tindep}
\end{eqnarray}
It is evident that, 
although $U_{2N}$ itself is not an observable,
$U_{2N}$ is always faithful to the phase shift by construction.
Another important property is that, 
while $U_{2N}$ could be energy-dependent in general,
it was proven that one can construct $U_{2N}$
so that it becomes energy-independent~\cite{Aoki:2008hh,Aoki:2009ji,Aoki:2012tk,Aoki:E-indep:inelastic}.

In lattice QCD, 
the NBS wave function of the ground state, $\phi_{2N}^0$, 
can be extracted from the four-point correlator as
\begin{eqnarray}
\label{eq:4pt_2N}
G_{2N} (\vec{r},t-t_0)
&\equiv& 
\frac{1}{L^3}
\sum_{\vec{R}}
\langle 0 |
          (N(\vec{R}+\vec{r}) N (\vec{R}))(t)\
\overline{(N' N')}(t_0)
| 0 \rangle , \\
&\xrightarrow[t \gg t_0]{}& A_{2N}^0 \phi_{2N}^0 e^{-E_{2N}^0(t-t_0)} , 
\quad
A_{2N}^0 = \langle E_{2N}^0 | \overline{(N' N')} | 0 \rangle , \\
\label{eq:NBS_2N}
\phi_{2N}^0(\vec{r}) &\equiv& \langle 0 | N(\vec{r}) N(\vec{0}) | E_{2N}^0\rangle ,
\end{eqnarray}
where 
$E_{2N}^0$ denotes the energy of the ground state,
$N$ ($N'$) the nucleon operator in the sink (source).
In the practical lattice calculation, however, it is notoriously difficult
to achieve the ground state saturation~\cite{HALQCD:2012aa}. 
This is because (i) the energy splitting between the ground state and excited states
are getting smaller for larger lattice volume
and (ii) signal to noise ratio (S/N) in lattice Monte Carlo simulation
is ruined exponentially with increasing $t$.
S/N also becomes exponentially worse for larger mass number in the system,
and/or for lighter pion mass on the lattice.

In order to overcome this problem,
it is recently proposed~\cite{HALQCD:2012aa} to consider the time-dependent Schr\"odinger equation,
\begin{eqnarray}
-\frac{\nabla^2}{2\mu} \psi_{2N}(\vec{r},t)
+ \int d\vec{r}' U_{2N}(\vec{r},\vec{r}') \psi_{2N}(\vec{r}',t)
= - \frac{\partial}{\partial t} \psi_{2N}(\vec{r},t) ,
\label{eq:Sch_2N:tdep}
\end{eqnarray}
where $\psi_{2N}$ is imaginary-time NBS wave function defined by
\begin{eqnarray}
\psi_{2N}(\vec{r},t) \equiv G_{2N} (\vec{r},t) / e^{-2m_N t} .
\end{eqnarray}
What is noteworthy is that, 
thanks to the energy-independence of $U_{2N}$,
Eq.~(\ref{eq:Sch_2N:tdep}) holds not only for the ground state but also for excited states simultaneously.
Therefore, the ground state saturation is not required,
which is a significant advantage of the potential approach to multi-baryon systems in lattice QCD.

In the practical calculation,
we perform the derivative expansion for the non-locality of the potential, 
%
%
$
U_{2N}(\vec{r},\vec{r}') =
\left[ V_C(r) + V_T(r) S_{12} + V_{LS}(r) \vec{L}\cdot \vec{S} + {\cal O}(\nabla^2) \right]
\delta(\vec{r}-\vec{r}') ,
$
%
%
where $V_C$, $V_T$ and $V_{LS}$ are the central, tensor and spin-orbit potentials, respectively.
In Ref.~\cite{Murano:2011nz}, the validity of this expansion is examined,
and it is shown that the leading terms, $V_C$ and $V_T$, dominate the potential at low energies.

We now extend the method to the 3N system.
We consider the imaginary-time NBS wave function of the 3N,
$\psi_{3N}(\vec{r},\vec{\rho},t)$,
defined by the six-point correlator as
\begin{eqnarray}
\label{eq:6pt_3N}
G_{3N} (\vec{r},\vec{\rho},t-t_0) 
&\equiv& 
\frac{1}{L^3}
\sum_{\vec{R}}
\langle 0 |
          (N(\vec{x}_1) N(\vec{x}_2) N (\vec{x}_3))(t) \
\overline{(N'       N'        N')}(t_0)
| 0 \rangle , \\
\label{eq:NBS_3N}
\psi_{3N}(\vec{r},\vec{\rho},t-t_0) &\equiv& G_{3N} (\vec{r},\vec{\rho},t-t_0) / e^{-3m_N (t-t_0)}
\end{eqnarray}
where
$\vec{R} \equiv ( \vec{x}_1 + \vec{x}_2 + \vec{x}_3 )/3$,
$\vec{r} \equiv \vec{x}_1 - \vec{x}_2$, 
$\vec{\rho} \equiv \vec{x}_3 - (\vec{x}_1 + \vec{x}_2)/2$
are the Jacobi coordinates.
With the derivative expansion of the potentials,
the NBS wave function can be converted to the potentials
through the following 
Schr\"odinger equation,
\begin{eqnarray}
%
\biggl[ 
- \frac{1}{2\mu_r} \nabla^2_{r} - \frac{1}{2\mu_\rho} \nabla^2_{\rho} 
+ \sum_{i<j} V_{2N} (\vec{r}_{ij})
+ V_{3NF} (\vec{r}, \vec{\rho})
\biggr] \psi_{3N}(\vec{r}, \vec{\rho},t)
= - \frac{\partial}{\partial t} \psi_{3N}(\vec{r}, \vec{\rho},t) , \ \ \ \ 
\label{eq:Sch_3N}
\end{eqnarray}
where
$V_{2N}(\vec{r}_{ij})$ with $\vec{r}_{ij} \equiv \vec{x}_i - \vec{x}_j$
denotes the 2NF between $(i,j)$-pair,
$V_{3NF}(\vec{r},\vec{\rho})$ the 3NF,
$\mu_r = m_N/2$, $\mu_\rho = 2m_N/3$ the reduced masses.
If we calculate 
$\psi_{3N}(\vec{r}, \vec{\rho},t)$,
and if all $V_{2N}(\vec{r}_{ij})$ are obtained
by (separate) lattice calculations for genuine 2N systems,
we can extract $V_{3NF}(\vec{r},\vec{\rho})$ through Eq.~(\ref{eq:Sch_3N}).

One of the greatest challenges 
in the study of multi-baryon systems on the lattice 
is that the computational cost of the correlators is exceptionally enormous.
Actually, for larger mass number in the system,
the cost for color/spinor contractions grows exponentially,
and the cost for Wick contractions grows factorially:
The total cost is the multiplication of both costs.
In order to meet this challenge,
we recently develop a novel algorithm in which 
color/spinor contractions and Wick contractions are considered 
simultaneously in a unified contraction index list~\cite{Doi:2012xd}.
In this ``unified contraction algorithm,''
redundancies in the original contractions are eliminated,
and a significant reduction in the computational cost is achieved,
e.g., 
by a factor of 192 for $^3$H and $^3$He correlators, and 
a factor of 20736 for the $^4$He correlator.
We note that,
since a potential is independent of the choice of a nucleon operator at the source,
we here employ the non-relativistic limit operator at the source
to maximize the gain by the unified contraction algorithm.
For the nucleon operator at the sink, which defines the NBS wave function
(and correspondingly, the ``scheme'' of the potential~\cite{Aoki:2009ji,Aoki:2011ep,Aoki:2012tk}),
we employ the standard nucleon operator,
$N_{std} \equiv \epsilon_{abc}(q_a^T C \gamma_5 q_b) q_c$,
for both of 2NF and 3NF,
so that they are determined on the same footing.

%

In our first exploratory study of 3NF,
we restrict the geometry of the 3N.
More specifically, we consider the ``linear setup''with $\vec{\rho}=\vec{0}$,
with which 3N are aligned linearly with equal spacings of 
$r_2 \equiv |\vec{r}|/2$.
%
In this setup,
the third nucleon is attached
to $(1,2)$-nucleon pair with only S-wave.
Considering the total 3N quantum numbers of 
$(I, J^P)=(1/2,1/2^+)$,
the triton channel, 
the wave function can be completely spanned by
only three bases, which can be labeled
by the quantum numbers of $(1,2)$-pair as
$^1S_0$, $^3S_1$, $^3D_1$.
Therefore, the Schr\"odinger equation
leads to 
the $3\times 3$ coupled channel equations
with the bases of 
$\psi_{^1S_0}$, $\psi_{^3S_1}$, $\psi_{^3D_1}$.
The reduction of the dimension of bases 
is expected to improve the S/N as well.
It is worth mentioning that
considering the linear setup is not an approximation:
Among various geometric components of 
the wave function in the triton channel,
we calculate the (exact) linear setup component
as
a convenient choice to study 3NF.
While we can access only a part of 3NF from it,
we plan to extend the calculation to more general geometries
step by step,
toward the complete determination of the full 3NF.

We consider the identification of genuine 3NF.
It is a nontrivial work:
Although both of parity-even and parity-odd 2NF
are required to subtract 2NF part in Eq.~(\ref{eq:Sch_3N}),
parity-odd 2NF have not been obtained yet in lattice QCD.
(See, however, our recent progress~\cite{Murano:2011aa}.)
In order to resolve this issue,
we consider the following channel,
\begin{eqnarray}
\psi_S \equiv
\frac{1}{\sqrt{6}}
\Big[
-   \Pu \Nu \Nd + \Pu \Nd \Nu               
                - \Nu \Nd \Pu + \Nd \Nu \Pu 
+   \Nu \Pu \Nd               - \Nd \Pu \Nu
\Big]  ,
\label{eq:psi_S}
\end{eqnarray}
which is anti-symmetric
in spin/isospin spaces 
for any 2N-pair.
Combined with the Pauli-principle,
it is automatically guaranteed that
any 2N-pair couples with even parity only.
Therefore, we can extract 3NF unambiguously 
using only parity-even 2NF.
Note that no assumption on the choice of 3D-configuration of $\vec{r}$, $\vec{\rho}$
is imposed in this argument,
and we thus can take advantage of this feature
for future 3NF calculations with various setup of 3D-geometries.

\vspace*{-4mm}
\section{Lattice QCD setup and Numerical results}
\label{sec:results}

We employ
$N_f=2$ dynamical 
configurations
with mean field improved clover fermion 
and 
RG-improved
gauge action
generated by CP-PACS Collaboration~\cite{Ali Khan:2001tx}.
We use
598 configurations at
$\beta=1.95$ and
the lattice spacing of
$a^{-1} = 1.269(14)$ GeV,
and 
the lattice size of $V = L^3 \times T = 16^3\times 32$
corresponds to
(2.5 fm)$^3$ box in physical spacial size.
For $u$, $d$ quark masses, 
we take the hopping parameter at the unitary point
as
$\kappa_{ud} = 0.13750$,
which corresponds to
$m_\pi = 1.13$ GeV, 
$m_N = 2.15$ GeV and
$m_\Delta = 2.31$ GeV.
Lattice simulations are performed
at eleven physical points of the distance $r_2$. 
We use the wall quark source with Coulomb gauge fixing.
In order to enhance the statistics,
we perform the measurement 
on 32 wall sources using different time slices,
and the forward and backward propagations are averaged.
The results from both of 
total angular momentum $J_z=\pm 1/2$ 
are averaged as well.
For the sink time, we calculate for a wide range of $5 \leq (t-t_0)/a \leq 10$.
We evaluate Eq.~(\ref{eq:Sch_3N}) at $t$ being integer or half-integer,
and we adopt the symmetric difference on the lattice for the time derivative.
In the case of $t$ of integer,
$\psi_{3N}(\bec{r},\bec{\rho},t)$ is obtained directly on the lattice,
while $\frac{\partial}{\partial t} \psi_{3N}(\bec{r},\bec{\rho},t)$ is obtained
from $\psi_{3N}(\bec{r},\bec{\rho},t \pm 1)$.
In the case of $t$ of half-integer,
both of $\psi_{3N}(\bec{r},\bec{\rho},t)$ and $\frac{\partial}{\partial t} \psi_{3N}(\bec{r},\bec{\rho},t)$
are evaluated from $\psi_{3N}(\bec{r},\bec{\rho},t \pm 1/2)$.

%
%

\begin{figure}[t]
\begin{minipage}{0.48\textwidth}
\begin{center}
\includegraphics[width=0.95\textwidth]{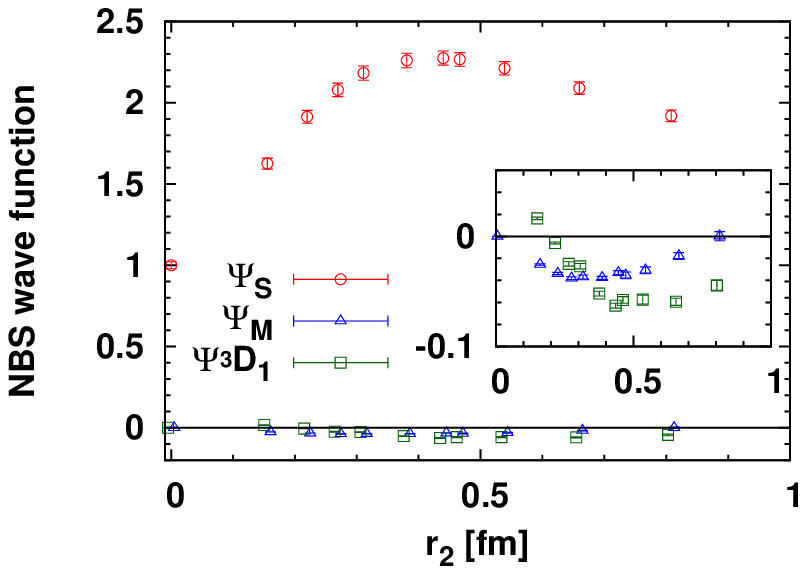}
\caption{
\label{fig:wf}
3N wave functions in the triton channel
at $(t-t_0)/a=8$.
Circle (red), triangle (blue), square (green) points denote
$\psi_S$, $\psi_M$, $\psi_{\,^3\!D_1}$, respectively.
$r_2$ 
is
the distance between the center and edge
in the linear setup.
}
\end{center}
\end{minipage}
\hfill
\begin{minipage}{0.48\textwidth}
\begin{center}
\includegraphics[width=0.95\textwidth]{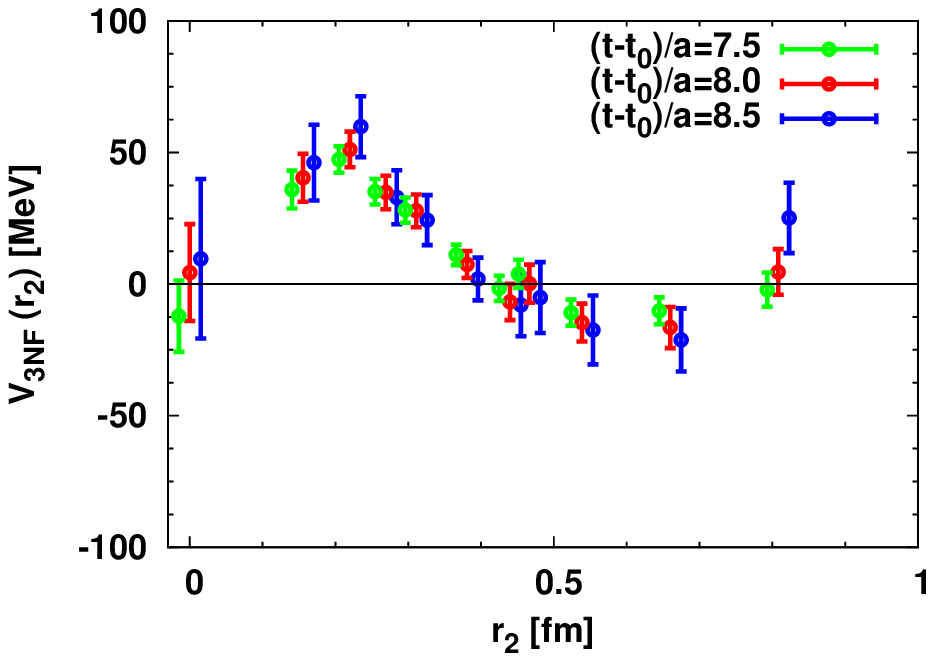}
\caption{
\label{fig:3N}
The effective scalar-isoscalar 3NF 
in the triton channel with the linear setup.
Green, red and blue points (with offset for visibility)
are
obtained at $(t-t_0)/a = 7.5$, 8.0 and 8.5, respectively.
}
\end{center}
\end{minipage}
\end{figure}

%
In Fig.~\ref{fig:wf},
we plot
the radial part of each wave function of
$\psi_S = ( - \psi_{^1S_0} + \psi_{^3S_1} )/\sqrt{2}$,
$\psi_M \equiv ( \psi_{^1S_0} + \psi_{^3S_1} )/\sqrt{2}$
and
$\psi_{^3D_1}$ 
obtained at $(t-t_0)/a = 8$.
Here, we normalize the wave functions
by the central value of $\psi_S(r_2=0)$.
What is noteworthy is that
the wave functions are obtained with good precision,
which is quite nontrivial for the 3N system.
We observe that 
$\psi_S$ overwhelms other wave functions.
This indicates 
higher partial wave components 
are strongly suppressed,
and the effect of the next leading order in the derivative expansion,
spin-orbit forces,
is suppressed in this lattice setup.

We determine 3NF
by subtracting 2NF from total potentials in the 3N system.
Since we have only one channel (Eq.~(\ref{eq:psi_S})) 
free from parity-odd 2NF,
we can determine one type of 3NF.
In this work,
3NF are
effectively represented in 
a scalar-isoscalar functional form,
which is 
often employed for
phenomenological short-range 3NF~\cite{Pieper:2001ap}.

In Fig.~\ref{fig:3N}, we plot the preliminary results
for the effective scalar-isoscalar 3NF at $(t-t_0)/a = 7.5, 8.0, 8.5$.
The results from different sink times are 
consistent with each other within statistical fluctuations.
This indicates that
contaminations from excited states above inelastic threshold 
and/or the higher order terms in the velocity expansion of the potential
are marginal.
%
%
%
%
Fig.~\ref{fig:3N} shows that
3NF are small at the long distance region of $r_2$.
This is in accordance with the suppression
of 2$\pi$E-3NF by the heavy pion.
At the short distance region, 
however,
an indication of repulsive 3NF is observed.
We recall that a repulsive short-range 3NF 
is phenomenologically required 
to explain the properties of high density matter.
Since multi-meson exchanges are strongly suppressed 
by the large quark mass, 
the origin of this short-range 3NF may be attributed to the 
quark and gluon dynamics directly.
In fact, we recall that the short-range repulsive (or attractive) cores
in the generalized two-baryon potentials 
are systematically 
calculated in lattice QCD in the flavor SU(3) limit, 
and the results are found to be well explained 
from the viewpoint of the Pauli exclusion principle in the quark level~\cite{Inoue:2010hs,Inoue:2011ai}.
In this context, 
it is intuitive to expect that the 3N system is subject to extra Pauli repulsion effect,
which could be an origin of the observed short-range repulsive 3NF.
Further investigation along this line is certainly an interesting subject in future.
It is also of interest that 
the analyses with operator product expansion~\cite{Aoki:OPE} show that
3NF has a repulsive core at short distance.

Evaluation of systematic errors are in progress.
In particular, 
the effect of the discretization error is important to be investigated,
since the nontrivial results are obtained at short distance.
For this purpose, an explicit lattice simulation with a finer lattice
is currently underway.
Quark mass dependence of 3NF 
is certainly 
an important issue as well, 
since the lattice simulations are carried out 
only at single large quark mass.
In the case of 2NF,
short-range cores have the enhanced strength
and broaden range by decreasing the quark mass%
~\cite{Aoki:2009ji}.
We, therefore, would expect a significant quark mass dependence
exist in short-range 3NF as well.
In addition,
long-range 2$\pi$E-3NF will emerge 
at lighter quark masses, in particular, at the physical point.
Quantitative investigation by
lattice simulations with lighter quark masses
are currently underway.

\vspace*{-3mm}
\section{Conclusions and Outlook}
\label{sec:summary}

We have explored three-nucleon forces (3NF)
in lattice QCD,
utilizing the imaginary-time Nambu-Bethe-Salpeter wave function.
The enormous computational cost is drastically reduced by
the newly-developed ``unified contraction algorithm.''
Using $N_f=2$ dynamical clover fermion configurations
at $a = 0.156$ fm, $V = 16^3\times 32$ and $m_\pi= 1.13$ GeV,
we have studied 3NF in which three nucleons are aligned linearly with an equal spacing. 
Repulsive 3NF have been found at short distance in the triton channel.

Currently, nuclear forces in lattice QCD are determined at rather heavy quark masses,
which is considered to be a largest source of systematic errors.
However, 
thanks to the significant theoretical and hardware development,
it becomes possible to perform lattice simulations at the physical quark mass point.
Generation of physical point gauge configurations
with large volume has started,
and nuclear forces will be subsequently studied~\cite{field5}.
While there remain various challenges,
it is becoming 
within reach to determine realistic nuclear forces including few-baryon forces 
from first-principles lattice simulations,
which will play an ultimate role in nuclear physics and astrophysics. 

\begin{acknowledgements}
We thank authors and maintainers of CPS++\cite{CPS}.
We also thank  
CP-PACS Collaboration
and ILDG/JLDG~\cite{conf:ildg/jldg} for providing gauge configurations.
The numerical simulations have been performed on 
Blue Gene/L, Blue Gene/Q and SR16000 at KEK,
SR16000 at YITP in Kyoto University, 
T2K at University of Tsukuba
and
T2K and FX10 at Tokyo University.
This research is supported in part by 
MEXT Grant-in-Aid for Young Scientists (B) (24740146),
Scientific Research on Innovative Areas (No.2004: 20105001, 20105003),
the Large Scale Simulation Program of KEK, 
the collaborative interdisciplinary program at T2K-Tsukuba, 
and SPIRE (Strategic Program for Innovative REsearch).
\end{acknowledgements}



\end{document}